\documentclass[sn-mathphys,Numbered]{sn-jnl}

\usepackage{graphicx}%
\usepackage{multirow}%
\usepackage{amsmath,amssymb,amsfonts}%
\usepackage{mathrsfs}%
\usepackage[title]{appendix}%
\usepackage{xcolor}%
\usepackage{textcomp}%
\usepackage{manyfoot}%
\usepackage{booktabs}%
\usepackage{listings}%

\graphicspath{ 
	{./Figures/} 
	{./Figures/1/}
}

\begin{document}

\title[Article Title]{The origin of the dark bands found in the Solar Spectra and its consequences}

\author{\fnm{Prachurjo} \sur{Dutta Roy}}

\abstract{This paper discusses the history of Fraunhofer’s puzzling discovery of the fixed lines in various spectra (most notably of the sun) and the implications of these spectral “imperfections”. Moreover, the developments in spectroscopy by Kirchhoff, Bunsen, et al. in the 19th century and its effects on our understanding of the atomic structure are discussed.}


\maketitle

\section{Introduction}\label{sec1}

In 1704, Newton showed in Opticks\footnote{Newton was not the first to do this in history, with several people, such as Della Porta, Grimaldi, Descartes, Boyle, et al. also having performed the same phenomena with water or a glass prism/slab \cite{crone}. Della Porta was also able to generate a blue-yellow spectrum considering the prism to be a darkening factor. See Fig. 3.6 in \cite[p.~43]{crone}.}, the now-popular method of dispersion using a prism. Newton used the prism along with a camera obscura (pinhole camera). He focused a single beam of light on to a prism and changed its angle. After a little movement, a band of colours appeared with red at the top with slight angular deviation (suggesting least refraction) and violet at the bottom with significant deviation (most refraction) \cite{new}. 

Newton hypothesised that sunlight or white light was polychromatic, i.e., composed of multiple colours and that each of these colours was differently refrangible. Newton observed that there were seven colours\footnote{Interestingly, some sources say he observed 5 (or 6) and added ‘orange’ and ‘indigo’ to make the spectrum in alignment with the ‘perfect seven’ philosophy. I saw in Opticks that the number went from 5 \cite[pp.~27-28]{new} to 7 \cite[p.~41]{new}, via experiments that were performed progressively in detail. His idea of 'blue' was close to the 'sky' or 'cyan' and he called the 'modern blue' as 'indigo'.} (VIBGYOR) in the spectrum, 2 more than the 5 that were known then.

\begin{figure}[h]
	\begin{center}
		\includegraphics[width=\textwidth]{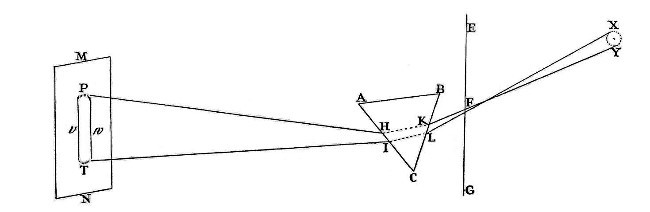}
		\caption{One of the first instances of Newton’s dispersion experiment in \cite{new}. The spectrum is generated at \textbf{PT} after passing through a narrow point at \textbf{F} and undergoing refraction at \textbf{KL} and \textbf{HI}. PC: Sir Isaac Newton Online}
	\end{center}		
\end{figure}

A relationship was also found between the red and violet ends of the light spectrum, 

\begin{equation}
	\mu_{violet}>\mu_{red}
	\label{eq1}
\end{equation}

Newton had tried to determine the velocities of colours within white light. And via Eq. 1 and the fact,

\begin{equation}
	v_v \mu_v=v_r \mu_r=c
	\label{eq2}
\end{equation}

This finding suggested that the speed of red was more than that of violet. This was one of the reasons why the corpuscular theory succeeded.

This, along with some other observations, was very instrumental in the development of the corpuscular theory. The phenomenon of dispersion became the driving force of this theory and the cause of its stronghold over the underdeveloped “undulation theory” \cite{crone}.

98 years later, in 1802, two primary developments are made, Thomas Young, a proponent of the wave theory, provides a rough idea of the frequencies and wavelengths of various colours in the visible spectrum in his Bakerian Lecture. He predicted a range of approximately 440 – 650 nm \cite[p.~39]{y02}. By then, the wave theory of light was slowly gaining traction by managing to explain the imperfections of the corpuscular theory.

\section{The Case of the Missing Rainbow}\label{sec2}

The second development was by William Wollaston, who used a novel method \cite{w02} to perform numerous experiments on various materials. He notes that unlike Newton’s seven-colour model, or the three-colour models described by others, the spectrum could not reducible by any means. He states however \cite[p.~378]{w02},

\begin{quote}
	"...by employing a very narrow pencil of light, 4 primary divisions of the prismatic spectrum may be seen, with a degree of distinctness that, I believe, has not been described nor observed before."
\end{quote}

\begin{figure}[h]
	\begin{center}
		\includegraphics[scale=0.85]{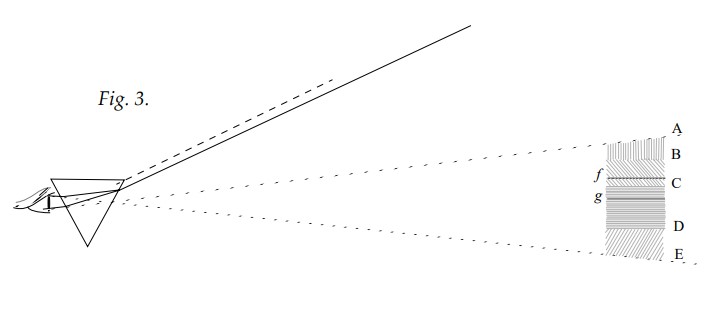}
		\caption{Wollaston’s prism experiment from which the inferences shown in points 1-5 are drawn. \textbf{AB}, \textbf{BC}, \textbf{CD}, \textbf{DE} represent the four colour segments as per him \cite{w02}.}
		\label{fig2}
	\end{center}		
\end{figure}

Interestingly, upon performing this experiment, he noticed a few properties of these lines as per Fig. \ref{fig2} \cite[p.~378]{w02},

\begin{enumerate}
	\item The boundary line \emph{A} on the red end seemed "somewhat confused", which he alludes to want of power in the eye to converge red light.
	\item \emph{A} distinct, dark line \emph{B} is observed between red and green.
	\item The limit of blue and green, line \emph{C} is not so clearly marked as the rest.
	\item On two sides of \emph{C}, there are two dark lines \emph{f} and \emph{g}, which he suggests, in an imperfect experiment, might be mistaken for the boundary of blue and green.
	\item \emph{D} and \emph{E} were found to represent the distinct blue-violet boundary and the violet end boundary.
\end{enumerate}

Wollaston suggests that this differs for each light source (having experimented on blue flame and blue electric light), giving rise to different divisions which he could not explain \cite{w02} \cite{j82}.

In 1814, Joseph Fraunhofer, performed a set of similar experiments independently. He wanted to measure the dispersion for every colour and every glass, however the indefinite limits of colours in the spectrum make this an uncertainty so great that it is useless. He does say that if there existed glasses or liquids that dispersed only one colour, the idea would be more plausible \cite{fps98}.

Fraunhofer observed that when white light was passed through a coloured liquid, the colour of the liquid appeared strongest in the spectrum. Recreating this with flames of alcohol, sulphur, etc., he observed a constant appearance of a sharply defined streak in the red-green intermediate region\footnote{If the region between red-green is to be thought of as yellow, then I am made to think if this streak (Wollaston Line B) was the first recorded instance of the modern D$_1$ – D$_2$ doublet.} of the light spectra \cite{fps98}. Another streak also appears in the green region, but is much feeble.

Experimenting with sunlight, he passed sunlight via a small slit\footnote{As per Fraunhofer, 15 seconds broad and 36 minutes high \cite[p.~4]{fps98}. Increasing the width to 40 seconds and 1 minute, made the fine and broad lines disappear respectively.} in a window for collimation. He passed it through a flint-glass prism, which was stood upon a theodolite at an incidence angle of $60^{\circ}$. His objective was to see if the same streak appeared in the red-green region. Instead, he observed (with a telescope) "almost countless" number of lines in the solar spectrum.

He noticed a few properties of these lines,

\begin{enumerate}

	\item The distances between the lines and their relations remained the same when the prism distance or the slit width were altered.
	
	\item The visibility of the lines was proportional to the size of the colour-image.

	\item The lines seemed to be invariant of the refracting object, i.e., the locations of the bands did not change with substance. 

	\item The strongest lines do not mark the limits of the colours as the same colour appear on both sides of most lines and any colour change is gradual and indistinguishable\footnote{A differentiation from Wollaston’s hypothesis is seen here.}.
	
	\item Similarly, it is impossible to set limits on both ends, although it is apparently easier at the red-end.

\end{enumerate}

Fraunhofer observed numerous lines, the most distinguishable of which he named from \emph{A} – \emph{I} \cite{fps98}\cite{sp10}.

\begin{figure}[ht]
	\begin{center}
		\includegraphics[width=\textwidth]{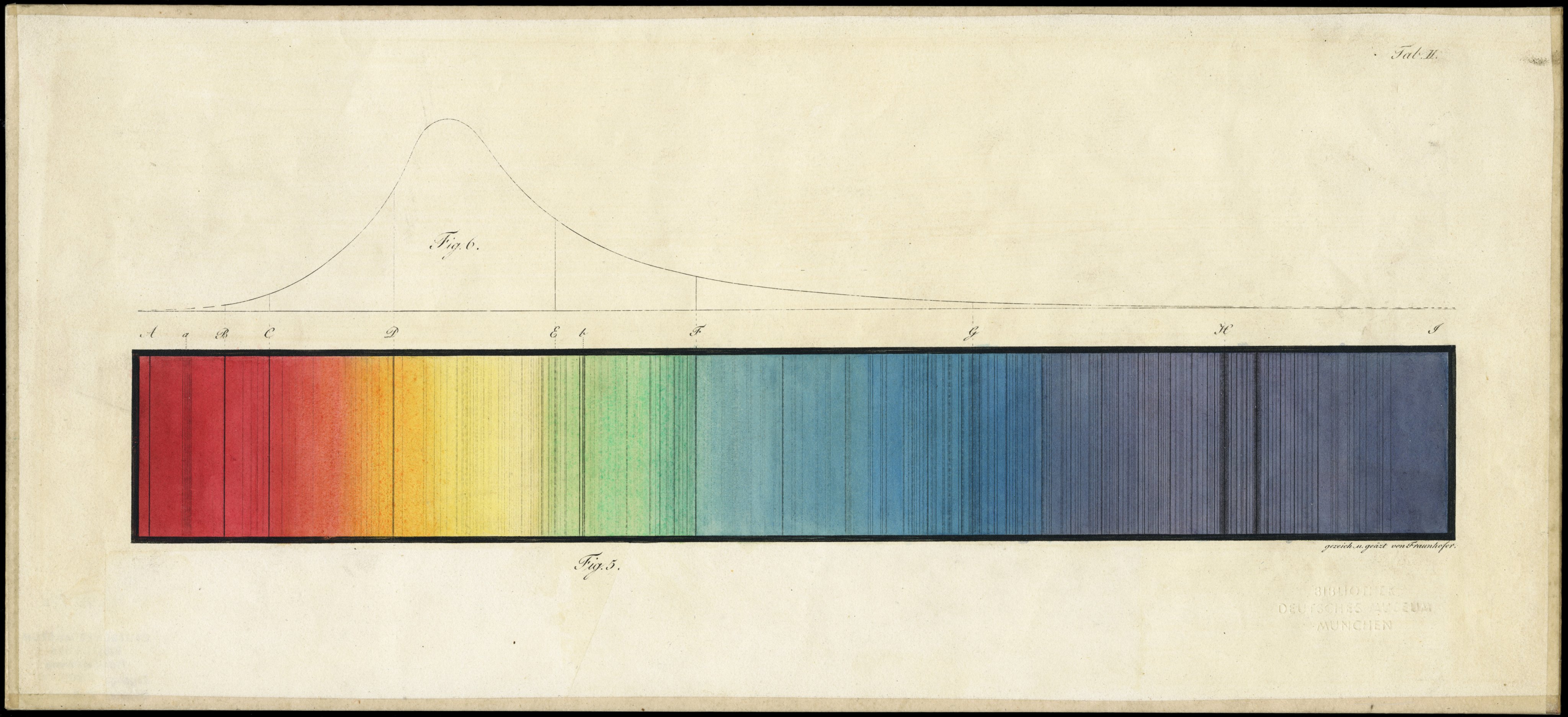}
		\caption{Fraunhofer’s spectrum, coloured version (1817). PC: Deutsches Museum}
		\label{fig3}
	\end{center}		
\end{figure}

The tabulated form (based on Fig. \ref{fig3} and the accounts of Fraunhofer in \cite{fps98}) of the total number of lines found by Fraunhofer is also given.

\begin{table}[ht]
\begin{tabular}{|l|r|}
\hline
\multicolumn{1}{|c|}{\textbf{REGION BETWEEN LINES}} & \multicolumn{1}{c|}{\textbf{NUMBER OF LINES IN THE REGION}} \\ \hline
A - B (Including a)                                 & Not documented\footnotemark[1]                                             \\ \hline
B - C                                               & 9                                                           \\ \hline
C - D                                               & 30                                                          \\ \hline
D - E                                               & 84                                                          \\ \hline
E - b                                               & 24                                                          \\ \hline
b - F                                               & 52                                                          \\ \hline
F - G                                               & 185                                                         \\ \hline
G - H                                               & 190                                                         \\ \hline
H - I                                               & Said to be equally numerous as G - H                        \\ \hline
TOTAL (Documented)                                  & 574                                                         \\ \hline
\end{tabular}
\caption{Detailed count of Fraunhofer Lines in the solar spectrum. A and I represent the spectrum extremities as per Fig. \ref{fig3}.}
\footnotetext[1]{Few lines are discussed by Fraunhofer in \cite[p.~5]{fps98}, where he suggests a sharp line at A itself, and at point a where many of the lines are heaped together forming a band.}
\label{table1}
\end{table}

These “missing 574 lines of the rainbow” seemed to be a wondrous perplexity for a long time, with people struggling to understand what they were. Wollaston saw them, but thought little of them. 

\section{Fingerprints in the Rainbow}\label{sec3}

Fraunhofer furthered his experiments by observing the spectra of various objects. He tested this out on Venus – where although the spectral ends were much feebler – and was able to see the \emph{D}, \emph{E}, \emph{b}, \emph{F} lines. He also distinguished the two \emph{b} lines – one weak and one strong. He was able to ascertain that Venus and sunlight had the same nature of light \cite{fps98}.

Similar experiments were carried on the star Sirius, whereupon he found three bands – one in green, two in blue. Electric light yielded few notable streaks in the green region and a feeble one in orange-yellow. Lamplight yielded two bands in the red-yellow region similar to the yellow doublet  (Fraunhofer lines D$_1$ and D$_2$). Hydrogen or alcohol combustion spectra show the reddish-green line in great detail, sulphur however, does not show so \cite{fps98}.

In the Max Planck Research magazine \cite{b21}, Thomas Bührke writes,

\begin{quote}
    "Spurred on by this discovery, he also found dark lines in the spectra of very bright stars – sometimes at the same positions but with different widths and intensities compared to the lines in the solar spectrum. It is precisely these differences that provide information about the composition and nature of each star."
\end{quote} 

\section{Deciphering the Lines}\label{sec4}

Building upon the then-emerging ‘undulatory theory’ of light, Anders Jonas Ångström uses Euler’s wave-based colour theory \cite{ped}, to put forward the concepts of absorption, interference and diffusion.

One of the key points behind the development of this theory was the Eulerian concept of each colour wave possessing its own particular wavelength and frequency, which determined the colour of the body. He further discusses the phenomena surrounding absorption and diffusion in various objects believing it to be fundamental to understanding any upcoming theories of light. In the paper, Ångström postulates that the electrical spectra\footnote{Generated by an electric spark which is then passed through a gas.} of different metals would be the best way to study any spectral phenomena \cite[p.~329]{ang}. Another idea put forward by him is that the electrical spectra would actually be the combination of two distinct spectra, namely, of the conductor itself and of the gas which acts as the spark medium \cite{ang}. 

He works with various elements such as lead (yellow and blue-violet), zinc (red and blue), etc. Another thing noted by Ångström is the appearance of the lines in clusters and also being common to one another in case of certain metals \cite{ang}. Extensive comparisons are made between the solar spectrum investigated by Wollaston and Fraunhofer and the electrical spectra which yield some common lines excepting a few notable ones such as the yellow doublet \emph{D} and the green Fraunhofer line \emph{b} \cite[p.~330]{ang}. Ångström also suggests that the doublet \emph{D} may correspond to one of the lines of bismuth, but further experiments showed that the various spectra lines of metals did appear very close to the doublet but was not exactly matching with the solar spectrum \cite[p.~332]{ang}. He also suggests that the ‘complete correspondence’ between the two spectra is seen to be very less due to the solar spectra originating from both the atmosphere and the sun.

Other than this, several other ideas are put forward by Ångström:

\begin{enumerate}
    \item Based on observations of Wheatstone, in a mix of two metals the spectra seen is of a combined one. The example given is that of the alloy Sn$_4$Pb, where the lines formed belong to both the metals. Changes in composition lead to very minimal shift toward the violet. \cite[p.~333]{ang}.
    \item Interference is ruled out as the reason behind the appearance of ‘peculiar lines’.
    \item The visible spectra generated by a material is due to its vibrations corresponding to the frequency of certain colours. 
    \item In case of electric spectra, the number of lines is directly proportional to the electricity strength.
\end{enumerate}

Further analyses led him to the fact that the dark lines generated by the solar spectra must be basically some sort of a reversion of the bright lines of the electric spectra \cite[p.~332]{ang}. Later on, he suggests that in cases of the gases, the spectral lines emitted were the same as the ones absorbed. This led him to suggest that a gas emits the same frequencies it absorbs.

He also suggests that studies on spark spectroscopy would lead to a solution on Doppler’s hypothesis on the colours of double stars being dependent on their velocity. Using a practical demonstration based on Wheatstone and Masson, Ångström suggests that the oscillation time and colour were in fact, independent of the velocity.

Ångström performs experiments on the specific gases\footnote{O$_2$, Carbonic Acid, H$_2$, NO$_2$, Carburetted Hydrogen (such as methane) and N$_2$.} in order to explain his ‘dual spectrum’ idea in case of spark spectra. All comparisons were made against an ‘air spectrum’. Oxygen spectrum showed an immediate absence of the four most prominent lines in the air spectrum and the appearance of several lines in the blue-violet fields. Carbonic acid showed a spectrum exactly similar to oxygen plus a few new lines which were seen in the air spectrum. The was attributed to the fact that the acid broke up to form CO$_2$ and O$_2$ \cite[p.~336]{ang}. Nitrous oxide showed an almost complete union of the air and oxygen spectra, although the explanation is said to be complicated if we consider the decomposition reactions of nitric oxide as explained by JJ Berzelius \cite[p.~337]{ang}. Nitrogen itself, showed that most of the lines in the air spectra were in fact caused by nitrogen and the air spectrum was not in fact a combustion of nitrogen and oxygen.

The most important contribution of \cite{ang} I believe was the studies on the spectra of Hydrogen and its compounds. The spectrum of Hydrogen was seen to have three notable lines (red, blue-green, extreme blue). Alcohol is said to have a similar spectrum with the red lines closer to the yellow. Carburetted hydrogen yielded the hydrogen spectrum and a notable line in the green region. Ångström suggests that spark decomposition of carburetted hydrogen into its constituent components may lead to these peculiar lines. In view of the additive nature of spectra, as discussed earlier in case of metals and also seen in the case of air, nitrogen and nitrogen dioxide; it can be said that the anomalous lines seen when compared to the spectra of carbon dioxide and oxygen, are that of carbon. It is worth noting that his analysis of coal did not show any peculiar lines \cite[p.~337]{ang}.

Ångström postulates that the entirety of the solar spectrum has the ability to possess chemical power. He shows the example of the blue-coloured hydrogen flame and more interestingly oxygen’s tendency to favour chemical reactions towards the violet end (actually in the UV range) of the spectrum with metals like zinc. This, as he suggests, “…proves at the same time that the oscillations of oxygen belong more particularly to the blue and violet portions of the spectrum” \cite[p.~340]{ang}.

In 1868, Ångström represented the wavelengths of the spectral lines as the seventh power of the millimetre (equivalent to 10$^{-10}$ m) \cite{ang69}. This later on became the standard unit to represent wavelengths\footnote{Now mostly deprecated. The nanometre (10$^{-9}$ m or nm) is used instead.}, named after him.

\section{Investigation of the Metallic Spectra}\label{sec5}

In the August 1860 section of the \emph{Philosophical Magazine}, the same volume where Kirchhoff published his theory on black bodies \cite{k60}, we see his and Bunsen’s work on trying to decipher these dark bands.

It was common knowledge by then that certain compounds tend to give off specific colours when heated. When this coloured light thus produced is analysed by a prism, various spectra exhibiting differently coloured bands or lines of light are seen \cite{kb60}.

Their experiments were limited to Group 1 and Group 2 elements\footnote{Li, Na, K (in texts as Ka), Ca, Sr, Ba were the known chemicals used as per \cite{kb60}.} and were discussed as a series of examples. The clarity of the bright lines was seen to be best at the highest flame temperature and least illuminating power and one such lamp\footnote{The lamp used to generate the flame was described by Bunsen and Roscoe in the \emph{Philosophical Transactions} in 1857. See \cite{br57} for further reading.} (now known as a Bunsen Burner) was used. The spectra generated was then compared against the solar spectra for comparison. All salts were purified to the highest extent possible and then treated with HCl to generate chloride salts for the experiment. 

\begin{figure}[ht]
	\begin{center}
		\includegraphics[width=\textwidth]{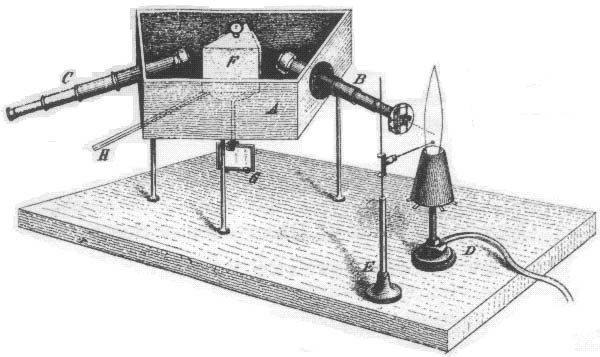}
		\caption{Kirchhoff and Bunsen’s spectrometer setup for analysing the spectra generated by chloride-salts of various elements \cite[p.~91]{kb60}.}
	\end{center}		
\end{figure}

Dry globules of the salts were then placed on a platinum wire held by a support (marked as \emph{E} in Figure ) and then brought to a flame, similar to modern flame test methods and after lengthy experiments by using various chemical flames (sulphur, CS$_2$, aqueous alcohol, H$_2$, etc.), observed the following \cite[p.~92]{kb60},

\begin{quote}
    “…the variety in the nature of the chemical processes occurring in the several flames, and the wide differences of temperature which these flames exhibit, produce no effect upon the position of the bright lines in the spectrum which are characteristic of each metal.”
\end{quote}

The first tests on sodium (more specifically, ordinary salt or sodium chloride) yielded an extraordinary coincidence! The sodium line Na $\alpha$ was coincident to the Fraunhofer \emph{D} Line and had a remarkably defined form and brightness. When Kirchhoff and Bunsen attempted to observe the sodium spectrum along with the solar spectrum, they noticed that the dark line became even darker \cite{w32}. Lithium yielded the lines Li $\alpha$, a bright red line and Li $\beta$, a weak yellow line. The reactions and conditions for the lithium flame test were described as exceeding “certainty and delicacy all methods hitherto known in analytical chemistry.” \cite[p.~96]{kb60}

Analysing potassium yielded mostly a continuous spectrum, apart from two distinct lines on the red and violet ends. These lines, Ka $\alpha$ and $\beta$, were observed to coincide with the Fraunhofer lines. The potassium-alpha line (Ka $\alpha$) was seen to be coinciding with the Fraunhofer line \emph{A}. Another indistinct line is also discussed, which appears when the light intensity is very high and coincides with Fraunhofer line \emph{B} \cite{kb60}.

Spectral analysis of alkaline earth metals started with strontium, which showed an absence of lines in the green region and eight lines that were notable: six red, one orange and one blue. Kirchhoff and Bunsen further brought down the important lines (based on position and intensity) to four: Sr $\alpha$ (orange), $\beta$ (red), $\gamma$ (red) and $\delta$ (blue). It is also suggested that Lithium is more distinguishable due to the more distinct appearance of Li $\alpha$ in the presence of Sr $\beta$. Calcium is distinguished by two bands: Ca $\alpha$ (green) and $\beta$ (orange-red) and has a spectrum different to that of Li, Na, K or Sr. Barium is said to have the “most complicated of the spectra of the alkalies [sic] and alkaline earths”. It is distinguished by two green lines: Ba $\alpha$ and $\beta$, along with a number of smaller lines\footnote{Other than the barium-gamma line, the other lines are mentioned only in the illustrations.} (Ba $\gamma$,$\delta$,$\eta$ (green) and $\epsilon$ (yellow), which was consistent with Sr $\alpha$).

\begin{figure}[ht]
	\begin{center}
		\includegraphics[width=\textwidth]{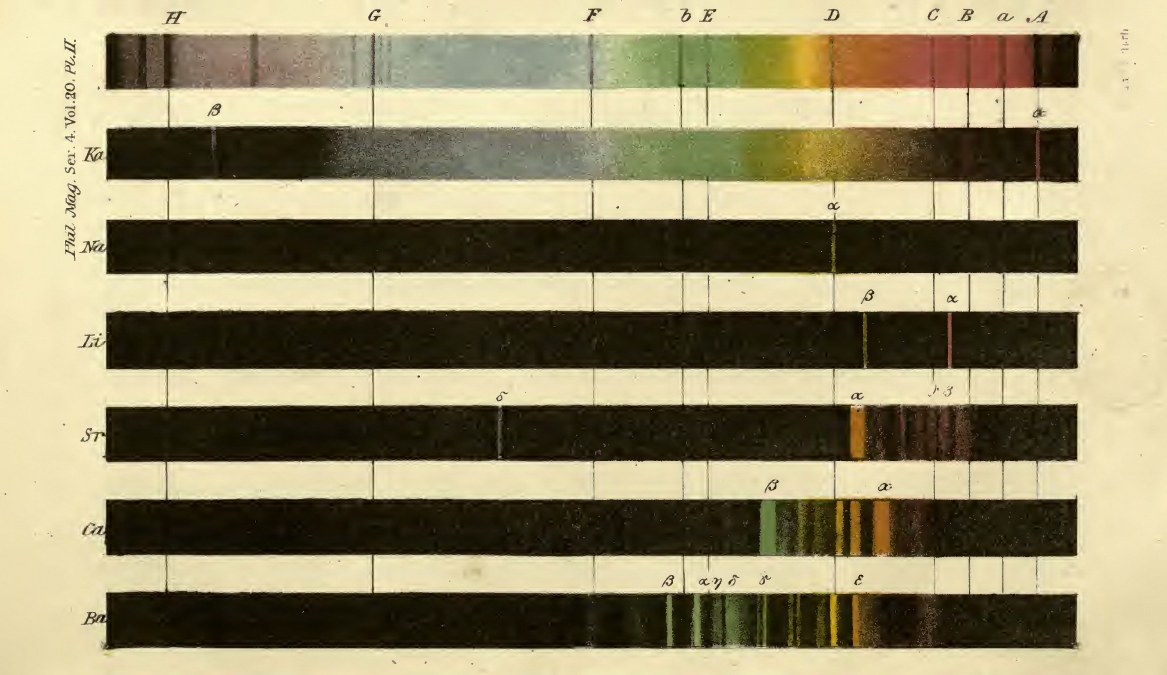}
		\caption{The spectrum of various alkalis and alkaline earth metals as given in \cite{kb60}.}
	\end{center}		
\end{figure}

Further analyses were carried out on various substances (cigar ash, mineral waters, seawater, Labradorite, etc.). This suggested that spectral analysis provided a simple and much easier method of identifying the components of a substance.

In \cite[p.~107]{kb60}, the authors write,

\begin{quote}
    “...The method of spectrum-analysis may also play a no less important part as a means of detecting new elementary substances; for if bodies should exist in nature so sparingly diffused that the analytical methods hitherto applicable have not succeeded in detecting, or separating them, it is very possible that their presence may be revealed by a simple examination of the spectra produced by their flames. We have had opportunity of satisfying ourselves that in reality such unknown elements exist. We believe that, relying upon unmistakeable results of the spectrum analysis, we are already justified in positively stating that, besides potassium, sodium, and lithium, the group of the alkaline metals contains a fourth member, which gives a spectrum as simple and characteristic as that of lithium—a metal which in our apparatus gives only two lines, namely a faint blue one, almost coincident with the strontium line Sr $\delta$, and a second blue one lying a little further towards the violet end of the spectrum, and rivalling the lithium line in brightness and distinctness of outline.”
\end{quote}

This prediction came about while analysing the mineral waters of Durkheim. The spectrum of the mineral water yielded the characteristic spectral lines of sodium, lithium, calcium and strontium \cite{w32}. After precipitation of the calcium, strontium and magnesium salts and treating the residue alcohol, nitric acid and separating the lithium components by ammonium carbonate, a ‘mother liquor’ was obtained whose spectrum showed two unexplained blue lines as written in the excerpt above. This was later revealed to be the element Caesium\footnote{Interestingly named after the Latin term for “bluish-grey”.}. Similar analyses in 1861 by Kirchhoff and Bunsen led to the spectrum of another element with distinct red lines, that came to be known as Rubidium\footnote{Following similar pattern to Caesium, named after the Latin term for “dark red”.} \cite{w32}.

Kirchhoff and Bunsen had also suggested that given the simplicity of the experiment, one could extend the scope of this activity to that of “an entirely untrodden field, stretching far beyond the limits of the earth, or even of our solar system.” \cite{kb60}

Another thing pointed out by them was that these spectra could get ‘reversed’\footnote{As already suggested by Ångström. See Section \ref{sec4}.}, i.e., the light bands would turn into dark bands and instead of a black screen, it would be visible on a spectrum (like that of Wollaston or Fraunhofer). The condition required for this reversal, as discussed in the black body paper, was the presence of a high-intensity light source (such as sunlight) originating from behind the source of the emission spectra. This is easily seen by the appearance of the sodium doublet when a sodium flame is kept in the presence of sunlight. This would suggest that if we were to see the spectrum yielded by the solar atmosphere, we would see the solar emission spectrum \cite[pp.~108-109]{kb60}. 
	
Kirchoff formulated the three laws of spectroscopy to represent the phenomena surrounding absorption and emission as follows, 
\begin{enumerate}
    \item Under high pressure, incandescent matter emits a continuous spectrum.
    \item A hot gas under low pressure emits bright lines, i.e., an emission spectrum.
    \item When a continuous spectrum is viewed through a low temperature, low density gas an absorption spectrum is produced.
\end{enumerate}

Kirchoff and Bunsen’s brilliant work laid the foundation in spectroscopy. As a result of their work, elements like thallium, indium, gallium, helium, ytterbium, holmium, thulium, samarium, neodymium, praseodymium, and lutecium were discovered \cite[p.~1424]{w32}.

At the end of their paper, they suggested that they would reserve further spectral analysis of terrestrial matter and the atmosphere of stars for a later paper.

\section{Movement of Heavenly Bodies}\label{sec6}

\begin{quote}
    “…news reached me of Kirchhoff's great discovery of the true nature and the chemical constitution of the sun from his interpretation of the Fraunhofer lines.

This news was to me like the coming upon a spring of water in a dry and thirsty land. Here at last presented itself the very order of work for which in an indefinite way I was looking — namely, to extend his novel methods of research upon the sun to the other heavenly bodies. A feeling as of inspiration seized me: I felt as if I had it now in my power to lift a veil which had never before been lifted; as if a key had been put into my hands which would unlock a door which had been regarded as for ever closed to man…” \cite[p.~46]{b11}
\end{quote}

William Huggins with the aid of his neighbour William Miller, had first investigated the chemical spectra in order to link stellar spectra to the ones produced by terrestrial elements \cite{h64}. Although an extension of Kirchhoff and Bunsen’s experiments, Huggins had successfully shown the relation of some more Fraunhofer lines to that of terrestrial elements. Apart from the sodium–\emph{D} relation by Kirchhoff and Bunsen, Huggins showed the relation of the red hydrogen line to Fraunhofer line \emph{C}. Links were also established between iron and lines \emph{E}, \emph{b$_3$} and \emph{G}, and calcium and \emph{H}\cite{h64}.

In the second paper \cite{hm64} on stellar spectra, various celestial spectra were studied in detail. Magnesium was linked to the remaining three b lines of Fraunhofer and hydrogen was linked to F. In studying the lunar spectra, it was found that there was no observable addition or disappearance of lines, except that of the reduced intensity of the solar lines. This was further proof of the fact that there was little to no atmosphere in the moon\footnote{This idea of determining atmospheres was already postulated by Kirchhoff and Bunsen in Section \ref{sec5}.}. Many other stars like Aldebaran, Sirius were observed and connections were made to the metallic spectra.

Huggins and Miller then discuss about the colours of stars and its effects on the resultant colours in the spectrum. The appearance of certain elements in the atmosphere of stars also makes them think on the distribution of the elements across parts of space. 

In 1868, Huggins writes that they had taken into account the possibility (during the experiments surrounding \cite{h64} and \cite{hm64}) of the stellar spectral lines no longer matching with their elemental counterparts, should the stars be in a motion relative to that of earth. The first set of experiments conducted during then produced no such displacement in the order of the difference between the two sodium lines \cite{h68}.

About 27 years prior, in order to explain the colours of double stars, Doppler had shown that the perception of a wave (light or sound) depends on the interval of time (frequencies) rather than the intrinsic strength and period \cite{d42}. He had underlined two cases\footnote{The notation used in \cite{d42} makes it quite confusing. Here $n$ =1⁄$f_o$  corresponds to the observed frequency and $x$=1⁄$f_e$ to the emitted frequency. Both $n$ and $x$ are units of time intervals}:

\begin{enumerate}
    \item When the observer approaches a stationary wave source with a speed $a_o$ and medium velocity $a$.
    \begin{equation}
        \frac{n}{x}=1+\frac{a_o}{a}
        \label{eq3}
    \end{equation}
    \item When the wave source approaches a stationary observer with a speed $a_s$ and medium velocity $a$.
    \begin{equation}
        \frac{n}{x}=\frac{a}{a-a_s}
        \label{eq4}
    \end{equation}
\end{enumerate}

Although not accepted during its time, the hypothesis slowly gained traction with Ballot’s successful experiments with soundwaves. As seen earlier, Ångström had failed to produce such an effect with light. Huggins had estimated the minimum required velocity to be about 315.4 km/s (196 mi/s) to produce the minimal shift of the spectral lines.
	
Maxwell had corresponded with Huggins on the matter and had shown \cite[pp.~532-533]{h68} that if a light source of frequency (emitted frequency) $n_e$ reaches the earth at some time $T$, and the system or the earth is in motion, then the light emanates $t$ seconds afterwards, reaching at a time $T'$. Since, $n_et$ vibrations had reached earth in the interval $t+T'-T$, the new frequency (observed frequency) comes as,

\begin{equation}
    n_o = n_e \frac{t}{t+T'-T}
    \label{eq5}
\end{equation}

Since, $vt=V(T'-T)$, where $v$ is the velocity of separation between the source and observer and $V$ the speed of light relative to earth, we can write Eq. \ref{eq5} as,

\begin{equation}
    \frac{n_e}{n_o}=1+\frac{T'-T}{t}=1+\frac{v}{V}
    \label{eq6}
\end{equation}

Now, $V=V_0-v_{earth}$, where $V_0$ is the speed of light in the aether. But we can approximate $V=V_0=c$ since it introduces a correction very small even for spectroscopic observations. So, we rewrite Eq. \ref{eq6} with the modern notation (a variation of Eq. \ref{eq3}),

\begin{equation}
    \frac{f_e}{f_o}=1+\frac{v}{c}
    \label{eq7}
\end{equation}

This can be further reduced to,

\begin{equation}
    \frac{f_o - f_e}{f_e}=-\frac{v}{c}
    \label{eq8}
\end{equation}

We can further see from this\footnote{The reason why $\Delta f$ carries a negative sign is because $f_e<f_o$.} \cite{ncert},

\begin{equation}
    \frac{\Delta f}{f_o} = - \frac{v}{c}
    \label{eq9}
\end{equation}

And also,

\begin{equation}
    \frac{\Delta \lambda}{\lambda_e}=\frac{\lambda_o - \lambda_e}{\lambda_e}=\frac{v}{c}
    \label{eq10}
\end{equation}

Here $v/c$ is taken as a value $z$ for ease of calculation. For earth, Bradley had already estimated this value to be 1/10210 in 1728 \cite{brad}. So, using any of the Eqs. \ref{eq8}-\ref{eq10}, we can calculate apparent ‘shifts’.

If the star moves away from the earth, then $z>0$ with an apparent shift towards the red regions, known as a \emph{redshift}. Similarly, movement toward the earth, would have a $z<0$ with a \emph{blueshift}. In case of the sodium doublets, the shift will be one-tenth (either way) of the difference of wavelengths of the two lines.

Armed with this new development, Huggins observed nebulae, notably the Great Nebulae in Orion\footnote{Now known as Messier 42 or M42.}, and saw that the nitrogen lines, especially the line at 500.8 nm appeared thinner in the nebulae than the spark spectra. Using Eq. \ref{eq10} with half the orbital velocity \cite{h68}, we can see $\Delta\lambda = 0.0249$ nm, a deviation too small to be measured. The experiments with nebulae seemed to be somewhat inconclusive, so he looks towards the stars.

In observing Sirius, Huggins noticed that the hydrogen line (Fraunhofer \emph{F}) in the spark and the nebula had a difference of about 0.109 nm suggesting a recessive motion relative to earth. The relative velocity calculated by Huggins is equal to 66.62 km/s (41.4 mi/s) \cite{h68}.

\begin{figure}[ht]
	\begin{center}
		\includegraphics[width=\textwidth]{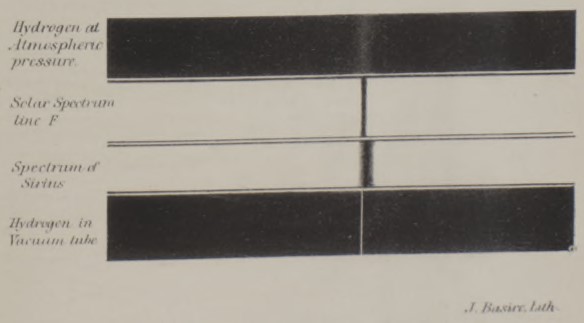}
		\caption{The spectrum of Sirius in \cite{h68}.}
	\end{center}		
\end{figure}

Similar experiments were also carried out on an eclipsed sun and a comet, named very originally, Comet II. Huggins spent the rest of his life on his newfound passion in spectroscopy with his wife Margaret Lindsay-Huggins. The pair contributed to a wide range of developments and discoveries surrounding the spectra of celestial objects. The story of this pair deserves its own paper which I hope to write about later.

\section{Into the 19th Century}\label{sec7}

In the last 15-20 years of the 19\textsuperscript{th} century, several advancements were made in the field of optics and optical instruments. Prisms were replaced by Rowland’s diffraction grating \cite{r82} and photography was invention. The existence of the ultraviolet and infrared was also now established. But most of the developments then were eclipsed by Maxwell’s formulation on the unification of electricity and magnetism, along with light.

In 1879, James Dewar and George Living studied the spectra of sodium and potassium \cite{dl79}. They found some empirical similarities in certain sodium doublets, like Group V (4983 Å, 4982 Å), Group VII (4667.2 Å, 4663.7 Å) and Group XI (4393 Å, 4390 Å) having a relation of nearly 1/15:1/16:1/17\footnote{An evidence that using wavenumbers would be ideal.}. Similar attempts were made with other doublets sodium and the quadruplets of potassium. Similar attempts were also made by Hartree, mostly along the lines of Dewar and Living. He found out that the difference in doublets or triplets had a constant difference of wavenumber. He had established periodic properties between certain elements as well \cite{h83}.

Balmer had examined the visible spectrum of hydrogen and had noticed that a particular wavelength factor was common to all four lines. This factor, $h$ was equal to 364.56 nm. Each of the related wavelength could thus be given by \cite{b85},

\begin{equation}
    \lambda = h \frac{n^2}{n^2 - 4}
    \label{eq11}
\end{equation}

Or in wavenumber form,

\begin{equation}
    \Bar{\nu} = \frac{1}{h} \left(1 - \frac{4}{n^2}\right) = \frac{4}{h} \left(\frac{1}{2^2}-\frac{1}{n^2}\right)
    \label{eq12}
\end{equation}

Plugging the first four values such that $n > 2$, i.e., 3, 4, 5, 6 give us the wavelengths: 656.208 nm, 486.08 nm, 434 nm and 410.13 nm, which was in very close agreement to the actual values.

Rydberg had found out that using wavenumbers would be simpler. In order to explain the relation of all spectra among alkali metals, he built a formula, 

\begin{equation}
    \Bar{\nu}=\Bar{\nu_0} - \frac{C_0}{(m + \mu)^2}
    \label{eq13}
\end{equation}

Where, $\Bar{\nu}$ is the spectral series limit, $m$ and $\mu$ are the line’s ordinal number and series specific constant respectively and $C_0$ is a universal constant. If we compare Eq. \ref{eq13} to Eq. \ref{eq12}, we see that $\mu=0$ for hydrogen and,

\begin{equation}
    C_0 = \frac{4}{h} \approx 10972.1 \mathrm{\ cm^{-1}}
    \label{eq14}
\end{equation}

This came to be known as the Rydberg Constant $R_\infty$\footnote{Current accepted value: 10973.7 cm$^{-1}$.}. Rydberg had also written groups of lines as ‘sharp’ or ‘diffused’ \cite{r90}. Principal referred to lower-ordinal ($m > 2$) doublets and fundamental to higher-ordinal ($m > 3$) triplets \cite{abgupta}. The general formula is given as,

\begin{equation}
    \Bar{\nu}=R_\infty \left(\frac{1}{n^2}-\frac{1}{m^2}\right)
    \label{eq15}
\end{equation}

In 1897, just one short paper after Wien’s paper on the emission spectra of blackbodies, Peter Zeeman had investigated the action of the magnetic field on light \cite{z97}. On surface, the results yielded negative results\footnote{Faraday had seen that there was some possible linkage of light with the electric field. But no observable changes were noted by him on alteration of the magnetic field \cite{fd33}.}. Zeeman observed the phenomena using a grating, focusing on the sodium doublet. On switching on the current, the \emph{D} lines were seen to widen and would revert back to their normal width on switching the current off. This was also seen with a different flame source and a different apparatus. Lithium showed anomalous behaviour in its red lines.

\begin{figure}[ht]
	\begin{center}
		\includegraphics[scale = 0.75]{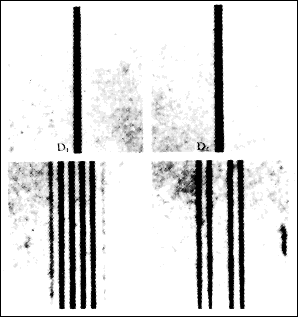}
		\caption{The splitting of the sodium doublet by action of the magnetic field. PC: chemteam.info}
	\end{center}		
\end{figure}

Atomic vibrations are now believed to be the cause of the spectral lines, and the magnetic field must alter it in some way, to represent this ‘widening’. Correspondences of Zeeman with Lorentz show that since, frequencies of the charged particles in an atom, must be related with those of the spectral lines. On an application of a magnetic field, the particles experience harmonic and Lorentz forces. The equation for the motion can be therefore given as \cite{z97} \cite{k97},

\begin{subequations}
    \begin{equation}
        ma_x = -kx + eB \frac{v_y}{c}
        \label{eq16a}
    \end{equation}
    \begin{equation}
        ma_y = -ky + eB \frac{v_x}{c}
        \label{eq16b}
    \end{equation}
    \begin{equation}
        ma_z = -kz
        \label{eq16c}
    \end{equation}
\end{subequations}

The value for $x$ and $y$ yield two solutions,

\begin{subequations}
    \begin{equation}
        x = a_1 (\cos \omega_1 t + p_1)
        \label{eq17a}
    \end{equation}
    \begin{equation}
        y = -a_1 (\sin \omega_1 t + p_1)
        \label{eq17b}
    \end{equation}
\end{subequations}

Or,

\begin{subequations}
    \begin{equation}
        x = a_2 (\cos \omega_2 t + p_2)
        \label{eq18a}
    \end{equation}
    \begin{equation}
        y = a_2 (\sin \omega_2 t + p_2)
        \label{eq18b}
    \end{equation}
\end{subequations}

We can then compute $\omega_1$ and $\omega_2$ as,

\begin{subequations}
    \begin{equation}
        \omega_1^2 - \frac{eB}{mc}\omega_1 = \omega_0^2
        \label{eq19a}
    \end{equation}
    \begin{equation}
        \omega_2^2 - \frac{eB}{mc}\omega_2 = \omega_0^2
        \label{eq19b}
    \end{equation}
\end{subequations}

This basically suggests that two ‘new’ lines will emerge from a base frequency $\omega_0$, appearing as two lines parallel to the field and three when perpendicular. Experimental observations by Zeeman later on, agreed with this \cite{zee97}.

Johannes Stark had shown a similar effect using electric field lines in 1913 \cite{s14}.

\section{Packets of Energy}\label{sec10} 

Soon after the resolving of the ‘ultraviolet catastrophe’ by Planck using quanta and Einstein’s explanation of the photelectric effect, physics took a new stride. ‘Quantum Physics’ was the talk of the town, having revolutionised the field of Optics and Thermodynamics. Lyman \cite{l06} and Paschen \cite{p08} had discovered spectral series of hydrogen beyond the visible ranges, using $n = 1$ and $n = 3$ respectively for the Rydberg formula (Eq. \ref{eq15}).

Walther Ritz had shown that the sum of two certain frequencies can give other frequencies in the spectrum \cite{abgupta}. This later on became a more generalised rule of Rydberg’s known as the Ritz or Rydberg-Ritz combination principle,

\begin{equation}
    \Bar{\nu}=A-\frac{R_\infty}{(n + \alpha + \beta(\alpha-\Bar{\nu}))^2}
    \label{eq20}
\end{equation}

Where, $A$ is the spectral limit and $\alpha$, $\beta$ correspond to constants.
	
After the famous Geiger-Marsden (gold-foil) experiment by Rutherford and his colleagues \cite{ncert} \cite{abgupta} \cite{r11}, the new planetary model of the atom became the best bet to explain the atomic structure. But the biggest opponent to it was spectroscopy. 

By Maxwell and Larmor, a charged particle undergoing circular motion will emit continuous radiation energy. Apart from the fact that the electron would eventually spiral into the nucleus, this radiation known as \emph{bremsstrahlung} (German: breaking radiation), would cause a continuous spectrum. But it was well established that the atomic spectra are discontinuous in nature. Another problem was the position of the electron orbits and the electron itself. This caused the start of the downfall of the Rutherford Model.

Niels Bohr was the one who finally solved the nature of these dark bands in 1913 \cite{ncert} \cite{abgupta} \cite{b13}. Based on the quantum theory, electrons could emit or absorb photons in packets of energy $\epsilon$  equivalent to,

\begin{equation}
    \epsilon = h\nu
    \label{eq21}
\end{equation}

Where, $\nu$ represents the frequency. Bohr extended this to atoms by suggesting specific stationary states of energy known as energy levels or stationary orbits. These were quantised such that the angular momentum of the n\textsuperscript{th} orbit would be,

\begin{equation}
    m_ev_nr_n = \frac{nh}{2\pi}
    \label{eq22}
\end{equation}

$n$ could be any integer.

An electron could ‘jump’ or ‘fall’ by absorbing or emitting a quanta. 

\begin{equation}
    E_1 + h\nu = E_2
    \label{eq23}
\end{equation}

Or,

\begin{equation}
    E_2 - h\nu = E_1
    \label{eq24}
\end{equation}

Or in a more general scenario,

\begin{equation}
    E_m - E_n = h(\nu_m - \nu_n) = h\Delta\nu_{nm}
    \label{eq25}
\end{equation}

Eq. \ref{eq25} is known as the \emph{Bohr Quantum Condition}.

The occurrence of absorption or emission lines are found to be coincident with the electron transition across two energy levels as energy is released in the form of a photon having a wavelength equal to that of the spectral line.

It was seen that the energy of the n\textsuperscript{th} orbit in the hydrogen atom was,

\begin{equation}
    E_n = \frac{1}{n^2}\frac{m_ee^4}{8 \varepsilon_0^2 h^2}
\end{equation}

Rewriting Eq. \ref{eq25} leads us to,

\begin{equation}
    h\nu = E_i - E_f = \frac{m_ee^4}{8 \varepsilon_0^2 h^2} \left(\frac{1}{n^2_f} - \frac{1}{n^2_i} \right)
\end{equation}

Rewriting $\nu=c/\lambda$,

\begin{equation}
   \frac{1}{\lambda} = \frac{m_ee^4}{8 \varepsilon_0^2 c h^3} \left(\frac{1}{n^2_f} - \frac{1}{n^2_i} \right) 
\end{equation}

If we compare this to Eq. \ref{eq15}, we see\footnote{In reality, $m$ is replaced by $\mu$, the reduced mass which is equal to $m_N/(m_N+m_e)$. For other atoms, a factor of $Z^2$, $Z$ being the atomic number, is introduced.},	

\begin{equation}
    R_\infty = \frac{m_ee^4}{8 \varepsilon_0^2 c h^3}
\end{equation}

Similarly, we can also show that the Ritz combination principle can also be described from Eqs. \ref{eq20} and \ref{eq25} for three orbits $a$, $b$ and $c$, given that $a>b>c$ as,

\begin{equation}
    h(\nu_a - \nu_c) = E_a - E_c
\end{equation}

Or,

\begin{equation}
    h(\nu_a - \nu_c )=(E_a-E_b )+(E_b-E_c)
\end{equation}

Which can be simplified to,

\begin{equation}
    \Delta\nu_{ac} = \Delta\nu_{ab} + \Delta\nu_{bc}
\end{equation}

\section{Conclusion}\label{sec13}

It is very interesting how a few dark lines in the solar spectra, which could have easily been mistaken for dispersive faults, ended up changing science in the way we know it today. Spectroscopic developments have been highlighted very less in compared to “mainstream optics, electromagnetism and atomic physics”. It is astonishing that Wollaston, Fraunhofer, Huggins, et al. are not names who we are familiar with, especially school students like us. These people have contributed to science in ways we could not have thought. 

I really think that the name of \cite{b21}, “Cosmic Detective Work” is a very apt name for what spectroscopists do. They investigate the missing rainbow, search for nature’s fundamental fingerprint. It is surprising that anyone from their windowed room can become such a detective, looking at the vast stars that are lightyears away from us and also understand how the atom works, using one single, simple method.

\backmatter

\bmhead{Acknowledgments}

I am deeply thankful to my mentor Dr. Ivy Dutta\footnote{Department of Physics, Delhi Public School, Ruby Park, Kolkata}, for guiding me and suggesting ideas on the material. 

A special mention goes to these books: \cite{crone} \cite{ncert} \cite{abgupta} and \cite{wich} for helping me in general with the historical and mathematical aspects of the paper and to the online library, Biodiversity Heritage Library, for having open access to a vast amount of historical texts.

\section*{Declarations}

\begin{itemize}
    \item Conflict of Interest - The corresponding author states that there is no conflict of interest.
    \item Data Availability - There is no data to be shared.
\end{itemize}

\begin{appendices}

\section{The Solar Spectra}\label{secA1}

I have provided here a table of the most important solar spectral lines for reference.

\begin{table}[h]
\resizebox{\columnwidth}{!}{
\begin{tabular}{|l|l|l|l|l|l|}
\hline
\textbf{WAVELENGTH   (nm)} & \textbf{DESG.} & \textbf{ELEMENT} & \textbf{WAVELENGTH   (nm)} & \textbf{DESG.} & \textbf{ELEMENT} \\ \hline
898.765 & y & Oxygen & 516.891 & b3 & Iron \\ \hline
822.696 & z & " & 516.733 & b4 & Magnesium \\ \hline
759.370 & A & " & 486.134 & F & Hydrogen-$\beta$   (Balmer) \\ \hline
686.719 & B & " & 434.047 & G’ & Hydrogen-$\gamma$ (Balmer) \\ \hline
656.281 & C & Hydrogen-$\alpha$   (Balmer) & 430.790 & G & Iron \\ \hline
627.661 & a & Oxygen & 430.774 & G & Calcium \\ \hline
589.592 & D1 & Sodium & 410.175 & h & Hydrogen-$\delta$   (Balmer) \\ \hline
588.995 & D2 & Sodium & 396.847 & H & Ionised Calcium \\ \hline
587.5618 & d & Helium & 393.366 & K & " \\ \hline
546.073 & e & Mercury & 382.044 & L & Iron \\ \hline
527.039 & E & Iron & 358.121 & N & " \\ \hline
518.362 & b1 & Magnesium & 336.112 & P & Ionised Titanium \\ \hline
517.270 & b2 & " & 302.108 & T & Iron \\ \hline
\end{tabular}}
\caption{Wavelengths of some major solar spectral lines.}
\label{table3}
\end{table}

\end{appendices}

\bibliography{sn-bibliography}


\begin{thebibliography}{38}
\ifx \bisbn   \undefined \def \bisbn  #1{ISBN #1}\fi
\ifx \binits  \undefined \def \binits#1{#1}\fi
\ifx \bauthor  \undefined \def \bauthor#1{#1}\fi
\ifx \batitle  \undefined \def \batitle#1{#1}\fi
\ifx \bjtitle  \undefined \def \bjtitle#1{#1}\fi
\ifx \bvolume  \undefined \def \bvolume#1{\textbf{#1}}\fi
\ifx \byear  \undefined \def \byear#1{#1}\fi
\ifx \bissue  \undefined \def \bissue#1{#1}\fi
\ifx \bfpage  \undefined \def \bfpage#1{#1}\fi
\ifx \blpage  \undefined \def \blpage #1{#1}\fi
\ifx \burl  \undefined \def \burl#1{\textsf{#1}}\fi
\ifx \doiurl  \undefined \def \doiurl#1{\url{https://doi.org/#1}}\fi
\ifx \betal  \undefined \def \betal{\textit{et al.}}\fi
\ifx \binstitute  \undefined \def \binstitute#1{#1}\fi
\ifx \binstitutionaled  \undefined \def \binstitutionaled#1{#1}\fi
\ifx \bctitle  \undefined \def \bctitle#1{#1}\fi
\ifx \beditor  \undefined \def \beditor#1{#1}\fi
\ifx \bpublisher  \undefined \def \bpublisher#1{#1}\fi
\ifx \bbtitle  \undefined \def \bbtitle#1{#1}\fi
\ifx \bedition  \undefined \def \bedition#1{#1}\fi
\ifx \bseriesno  \undefined \def \bseriesno#1{#1}\fi
\ifx \blocation  \undefined \def \blocation#1{#1}\fi
\ifx \bsertitle  \undefined \def \bsertitle#1{#1}\fi
\ifx \bsnm \undefined \def \bsnm#1{#1}\fi
\ifx \bsuffix \undefined \def \bsuffix#1{#1}\fi
\ifx \bparticle \undefined \def \bparticle#1{#1}\fi
\ifx \barticle \undefined \def \barticle#1{#1}\fi
\bibcommenthead
\ifx \bconfdate \undefined \def \bconfdate #1{#1}\fi
\ifx \botherref \undefined \def \botherref #1{#1}\fi
\ifx \url \undefined \def \url#1{\textsf{#1}}\fi
\ifx \bchapter \undefined \def \bchapter#1{#1}\fi
\ifx \bbook \undefined \def \bbook#1{#1}\fi
\ifx \bcomment \undefined \def \bcomment#1{#1}\fi
\ifx \oauthor \undefined \def \oauthor#1{#1}\fi
\ifx \citeauthoryear \undefined \def \citeauthoryear#1{#1}\fi
\ifx \endbibitem  \undefined \def \endbibitem {}\fi
\ifx \bconflocation  \undefined \def \bconflocation#1{#1}\fi
\ifx \arxivurl  \undefined \def \arxivurl#1{\textsf{#1}}\fi
\csname PreBibitemsHook\endcsname

\bibitem[\protect\citeauthoryear{Crone}{1999}]{crone}
\begin{bbook}
\bauthor{\bsnm{Crone}, \binits{R.A.}}:
\bbtitle{A History of Color: The Evolution of Theories of Light and Color}.
\bpublisher{Springer},
\blocation{Dordrecht}
(\byear{1999})
\end{bbook}
\endbibitem

\bibitem[\protect\citeauthoryear{Newton}{1730}]{new}
\begin{bbook}
\bauthor{\bsnm{Newton}, \binits{I.}}:
\bbtitle{Opticks Or, a Treatise of the Reflections, Refractions, Inflections, and Colours of Light}.
\bpublisher{William Innys},
\blocation{London}
(\byear{1730})
\end{bbook}
\endbibitem

\bibitem[\protect\citeauthoryear{Young}{1802}]{y02}
\begin{barticle}
\bauthor{\bsnm{Young}, \binits{T.}}:
\batitle{{"The Bakerian Lecture. On the theory of light and colours"}}.
\bjtitle{Philosophical Transactions of the Royal Society of London}
\bvolume{92},
\bfpage{12}--\blpage{48}
(\byear{1802})
\end{barticle}
\endbibitem

\bibitem[\protect\citeauthoryear{Wollaston}{1802}]{w02}
\begin{barticle}
\bauthor{\bsnm{Wollaston}, \binits{W.H.}}:
\batitle{{"A Method of examining refractive and dispersive Powers, by prismatic Reflection"}}.
\bjtitle{Philosophical Transactions of the Royal Society of London}
\bvolume{92},
\bfpage{365}--\blpage{380}
(\byear{1802})
\end{barticle}
\endbibitem

\bibitem[\protect\citeauthoryear{Johnson}{1882}]{j82}
\begin{barticle}
\bauthor{\bsnm{Johnson}, \binits{A.}}:
\batitle{{"Newton, Wollaston, and Fraunhofer's Lines"}}.
\bjtitle{Nature}
\bvolume{26}(\bissue{676}),
\bfpage{572}
(\byear{1882})
\end{barticle}
\endbibitem

\bibitem[\protect\citeauthoryear{Fraunhofer}{1898}]{fps98}
\begin{bbook}
\bauthor{\bsnm{Fraunhofer}, \binits{J.R.}}:
\bbtitle{Harper's Scientific Memoirs: Prismatic and Diffraction Spectra}
vol. \bseriesno{2}.
\bpublisher{Harper \& Brothers}, \blocation{???}
(\byear{1898})
\end{bbook}
\endbibitem

\bibitem[\protect\citeauthoryear{Suer and Pasachoff}{2010}]{sp10}
\begin{barticle}
\bauthor{\bsnm{Suer}, \binits{T.-A.}},
\bauthor{\bsnm{Pasachoff}, \binits{J.M.}}:
\batitle{{"The Origin and Diffusion of the H and K Notation"}}.
\bjtitle{Journal of Astronomical History and Heritage}
\bvolume{13}(\bissue{2}),
\bfpage{120}--\blpage{126}
(\byear{2010})
\end{barticle}
\endbibitem

\bibitem[\protect\citeauthoryear{Bührke}{2021}]{b21}
\begin{botherref}
\oauthor{\bsnm{Bührke}, \binits{T.}}:
{"Cosmic Detective Work"}.
Max Planck Research
(4),
66--72
(2021)
\end{botherref}
\endbibitem

\bibitem[\protect\citeauthoryear{Pedersen}{2008}]{ped}
\begin{barticle}
\bauthor{\bsnm{Pedersen}, \binits{K.M.}}:
\batitle{{"Leonhard Euler's Wave Theory of Light"}}.
\bjtitle{Perspectives on Science}
\bvolume{16}(\bissue{4}),
\bfpage{392}--\blpage{416}
(\byear{2008})
\end{barticle}
\endbibitem

\bibitem[\protect\citeauthoryear{Ångström}{1855}]{ang}
\begin{barticle}
\bauthor{\bsnm{Ångström}, \binits{A.J.}}:
\batitle{{"Optical researches"}}.
\bjtitle{The London, Edinburgh and Dublin Philosophical Magazine and Journal of Science}
\bvolume{9}(\bissue{60}),
\bfpage{327}--\blpage{342}
(\byear{1855})
\end{barticle}
\endbibitem

\bibitem[\protect\citeauthoryear{Ångström}{1869}]{ang69}
\begin{bbook}
\bauthor{\bsnm{Ångström}, \binits{A.J.}}:
\bbtitle{Recherches sur Le Spectre Solaire}.
\bpublisher{Ferdinand Dümmler}, \blocation{???}
(\byear{1869})
\end{bbook}
\endbibitem

\bibitem[\protect\citeauthoryear{Kirchhoff}{1860}]{k60}
\begin{barticle}
\bauthor{\bsnm{Kirchhoff}, \binits{G.R.}}:
\batitle{{“On the Relation between the Radiating and Absorbing Powers of different Bodies for Light and Heat,”}}.
\bjtitle{The London, Edinburgh, and Dublin Philosophical Magazine and Journal of Science}
\bvolume{20}(\bissue{130}),
\bfpage{1}--\blpage{21}
(\byear{1860})
\end{barticle}
\endbibitem

\bibitem[\protect\citeauthoryear{Kirchhoff and Bunsen}{1860}]{kb60}
\begin{barticle}
\bauthor{\bsnm{Kirchhoff}, \binits{G.R.}},
\bauthor{\bsnm{Bunsen}, \binits{R.W.}}:
\batitle{{“Chemical Analysis by Spectrum-observations,”}}.
\bjtitle{The London, Edinburgh and Dublin Philosophical Magazine and Journal of Science}
\bvolume{20}(\bissue{131}),
\bfpage{89}--\blpage{109}
(\byear{1860})
\end{barticle}
\endbibitem

\bibitem[\protect\citeauthoryear{Bunsen and Roscoe}{1857}]{br57}
\begin{barticle}
\bauthor{\bsnm{Bunsen}, \binits{R.W.}},
\bauthor{\bsnm{Roscoe}, \binits{H.E.}}:
\batitle{{“Photo-chemical researches.—Part I. Measurement of the chemical action of light,”}}.
\bjtitle{Philosophical Transactions of the Royal Society of London}
\bvolume{147},
\bfpage{355}--\blpage{380}
(\byear{1857})
\end{barticle}
\endbibitem

\bibitem[\protect\citeauthoryear{Weeks}{1932}]{w32}
\begin{barticle}
\bauthor{\bsnm{Weeks}, \binits{M.E.}}:
\batitle{{“The discovery of the elements XIII. Some spectroscopic discoveries,”}}.
\bjtitle{Journal of Chemical Education}
\bvolume{9}(\bissue{8}),
\bfpage{1413}--\blpage{1434}
(\byear{1932})
\end{barticle}
\endbibitem

\bibitem[\protect\citeauthoryear{Becker}{2011}]{b11}
\begin{bbook}
\bauthor{\bsnm{Becker}, \binits{B.J.}}:
\bbtitle{Unravelling Starlight}.
\bpublisher{Cambridge University Press}, \blocation{???}
(\byear{2011})
\end{bbook}
\endbibitem

\bibitem[\protect\citeauthoryear{Huggins}{1864}]{h64}
\begin{barticle}
\bauthor{\bsnm{Huggins}, \binits{W.}}:
\batitle{{“On the Spectra of some Chemical Elements,”}}.
\bjtitle{Philosophical Transactions of the Royal Society of London}
\bvolume{154},
\bfpage{139}--\blpage{160}
(\byear{1864})
\end{barticle}
\endbibitem

\bibitem[\protect\citeauthoryear{Huggins and Miller}{1864}]{hm64}
\begin{barticle}
\bauthor{\bsnm{Huggins}, \binits{W.}},
\bauthor{\bsnm{Miller}, \binits{W.A.}}:
\batitle{{“On the Spectra of some of the fixed stars,”}}.
\bjtitle{Philosophical Transactions of the Royal Society of London}
\bvolume{154},
\bfpage{413}--\blpage{435}
(\byear{1864})
\end{barticle}
\endbibitem

\bibitem[\protect\citeauthoryear{Huggins}{1868}]{h68}
\begin{barticle}
\bauthor{\bsnm{Huggins}, \binits{W.}}:
\batitle{{“Further observations on the spectra of some the stars and nebulæ, with an attempt to determine therefrom whether these bodies are moving towards or from the earth, also observations on the spectra of the sun and of comet II.,”}}.
\bjtitle{Philosophical Transactions of the Royal Society of London}
\bvolume{158},
\bfpage{529}--\blpage{564}
(\byear{1868})
\end{barticle}
\endbibitem

\bibitem[\protect\citeauthoryear{Doppler}{1842}]{d42}
\begin{barticle}
\bauthor{\bsnm{Doppler}, \binits{C.}}:
\batitle{{“Über das farbige Licht der Doppelsterne und einiger anderer Gestirne des Himmels,”}}.
\bjtitle{Proceedings of the Royal Bohemian Society of Sciences}
\bvolume{V}(\bissue{2}),
\bfpage{465}--\blpage{482}
(\byear{1842})
\end{barticle}
\endbibitem

\bibitem[\protect\citeauthoryear{NCERT}{2007}]{ncert}
\begin{bbook}
\bauthor{\bsnm{NCERT}}:
\bbtitle{Physics - XII (Part - II)},
\bedition{Unrationalised} edn.
\bpublisher{NCERT},
\blocation{New Delhi}
(\byear{2007})
\end{bbook}
\endbibitem

\bibitem[\protect\citeauthoryear{Bradley}{1728}]{brad}
\begin{barticle}
\bauthor{\bsnm{Bradley}, \binits{J.}}:
\batitle{{“Account of a new discovered motion of the fix'd stars,”}}.
\bjtitle{Philosophical Transactions of the Royal Society of London}
\bvolume{35}(\bissue{406}),
\bfpage{637}--\blpage{661}
(\byear{1728})
\end{barticle}
\endbibitem

\bibitem[\protect\citeauthoryear{Rowland}{1882}]{r82}
\begin{barticle}
\bauthor{\bsnm{Rowland}, \binits{H.A.}}:
\batitle{{“Preliminary notice of the results accomplished in the manufacture and theory of gratings for optical purposes,”}}.
\bjtitle{The London, Edinburgh, and Dublin Philosophical Magazine and Journal of Science}
\bvolume{13}(\bissue{84}),
\bfpage{469}--\blpage{474}
(\byear{1882})
\end{barticle}
\endbibitem

\bibitem[\protect\citeauthoryear{Dewar and Liveing}{1879}]{dl79}
\begin{barticle}
\bauthor{\bsnm{Dewar}, \binits{J.}},
\bauthor{\bsnm{Liveing}, \binits{G.D.}}:
\batitle{{“On the Spectra of Sodium and Potassium.,”}}.
\bjtitle{Proceedings of the Royal Society of London}
\bvolume{29}(\bissue{196-199}),
\bfpage{398}--\blpage{402}
(\byear{1879})
\end{barticle}
\endbibitem

\bibitem[\protect\citeauthoryear{Hartree}{1883}]{h83}
\begin{barticle}
\bauthor{\bsnm{Hartree}, \binits{W.N.}}:
\batitle{{“On Homologous Spectra,”}}.
\bjtitle{Journal of the Chemical Society, Transactions}
\bvolume{43},
\bfpage{390}--\blpage{400}
(\byear{1883})
\end{barticle}
\endbibitem

\bibitem[\protect\citeauthoryear{Balmer}{1885}]{b85}
\begin{barticle}
\bauthor{\bsnm{Balmer}, \binits{J.J.}}:
\batitle{{“Notiz über die Spectrallinien des Wasserstoffs”}}.
\bjtitle{Annalen der Physik}
\bvolume{261}(\bissue{5}),
\bfpage{80}--\blpage{87}
(\byear{1885})
\end{barticle}
\endbibitem

\bibitem[\protect\citeauthoryear{Rydberg}{1890}]{r90}
\begin{barticle}
\bauthor{\bsnm{Rydberg}, \binits{J.R.}}:
\batitle{{“On the structure of the line-spectra of the chemical elements,”}}.
\bjtitle{The London, Edinburgh, and Dublin Philosophical Magazine and Journal of Science}
\bvolume{29}(\bissue{179}),
\bfpage{331}--\blpage{337}
(\byear{1890})
\end{barticle}
\endbibitem

\bibitem[\protect\citeauthoryear{Gupta}{2011}]{abgupta}
\begin{bbook}
\bauthor{\bsnm{Gupta}, \binits{A.B.}}:
\bbtitle{Modern Atomic and Nuclear Physics}.
\bpublisher{Books and Allied}, \blocation{???}
(\byear{2011})
\end{bbook}
\endbibitem

\bibitem[\protect\citeauthoryear{Zeeman}{1897}]{z97}
\begin{barticle}
\bauthor{\bsnm{Zeeman}, \binits{P.}}:
\batitle{{“On the influence of magnetism on the nature of the light emitted by a substance,”}}.
\bjtitle{The London, Edinburgh and Dublin Philosophical Magazine and Journal of Science}
\bvolume{43}(\bissue{267}),
\bfpage{226}--\blpage{239}
(\byear{1897})
\end{barticle}
\endbibitem

\bibitem[\protect\citeauthoryear{Faraday}{1933}]{fd33}
\begin{bbook}
\bauthor{\bsnm{Faraday}, \binits{M.}}:
\bbtitle{Faraday's Diary: Being the Various Philosophical Notes of Experimental Investigation}
vol. \bseriesno{IV}.
\bpublisher{G. Bell and Sons Ltd},
\blocation{London}
(\byear{1933})
\end{bbook}
\endbibitem

\bibitem[\protect\citeauthoryear{Kox}{1997}]{k97}
\begin{barticle}
\bauthor{\bsnm{Kox}, \binits{A.J.}}:
\batitle{{“The discovery of the electron: II. The Zeeman effect,”}}.
\bjtitle{European Journal of Physics}
\bvolume{18},
\bfpage{139}--\blpage{144}
(\byear{1997})
\end{barticle}
\endbibitem

\bibitem[\protect\citeauthoryear{Zeeman}{1897}]{zee97}
\begin{barticle}
\bauthor{\bsnm{Zeeman}, \binits{P.}}:
\batitle{{“Doublets and triplets in the spectrum produced by external magnetic forces.—(II.),”}}.
\bjtitle{The London, Edinburgh, and Dublin Philosophical Magazine and Journal of Science}
\bvolume{44}(\bissue{268}),
\bfpage{255}--\blpage{259}
(\byear{1897})
\end{barticle}
\endbibitem

\bibitem[\protect\citeauthoryear{Stark}{1914}]{s14}
\begin{barticle}
\bauthor{\bsnm{Stark}, \binits{J.}}:
\batitle{{“Observations of the effect of the electric field on spectral lines I. Transverse effect,"}}.
\bjtitle{Annalen der Physik}
\bvolume{43},
\bfpage{965}--\blpage{983}
(\byear{1914})
\end{barticle}
\endbibitem

\bibitem[\protect\citeauthoryear{Lyman}{1906}]{l06}
\begin{botherref}
\oauthor{\bsnm{Lyman}, \binits{T.}}:
{“Preliminary Measurement of the Short Wave-Lengths Discovered by Schumann,”}.
Astrophysical Journal
\textbf{19}(263)
(1906)
\end{botherref}
\endbibitem

\bibitem[\protect\citeauthoryear{Paschen}{1908}]{p08}
\begin{barticle}
\bauthor{\bsnm{Paschen}, \binits{F.}}:
\batitle{{“Zur Kenntnis ultraroter Linienspektra. I. (Normalwellenlängen bis 27000 Å.-E.),”}}.
\bjtitle{Annalen der Physik}
\bvolume{332}(\bissue{13}),
\bfpage{537}--\blpage{570}
(\byear{1908})
\end{barticle}
\endbibitem

\bibitem[\protect\citeauthoryear{Rutherford}{1911}]{r11}
\begin{barticle}
\bauthor{\bsnm{Rutherford}, \binits{E.}}:
\batitle{{“The Scattering of $\alpha$ and $\beta$ Particles by Matter and the Structure of the Atom,”}}.
\bjtitle{The London, Edinburgh and Dublin Philosophical Magazine and Journal of Science}
\bvolume{21}(\bissue{125}),
\bfpage{669}--\blpage{688}
(\byear{1911})
\end{barticle}
\endbibitem

\bibitem[\protect\citeauthoryear{Bohr}{1913}]{b13}
\begin{barticle}
\bauthor{\bsnm{Bohr}, \binits{N.}}:
\batitle{{“On the constitution of atoms and molecules,”}}.
\bjtitle{The London, Edinburgh, and Dublin Philosophical Magazine and Journal of Science}
\bvolume{26}(\bissue{151}),
\bfpage{1}--\blpage{25}
(\byear{1913})
\end{barticle}
\endbibitem

\bibitem[\protect\citeauthoryear{Wichman}{2011}]{wich}
\begin{bbook}
\bauthor{\bsnm{Wichman}, \binits{E.H.}}:
\bbtitle{Berkeley Physics Course Volume 4: Quantum Physics (In SI Units)}.
\bpublisher{McGraw Hill}, \blocation{???}
(\byear{2011})
\end{bbook}
\endbibitem

\end{thebibliography}

\end{document}